\documentclass{IOS-Book-Article}

\usepackage{mathptmx}

\usepackage{amsmath}
\usepackage{graphicx}
\usepackage{subcaption}

\begin{document}

\pagestyle{headings}
\def\thepage{}

\begin{frontmatter}              

\title{Quantum Annealing and the Satisfiability Problem}


\author{\fnms{Kristen L} \snm{Pudenz}%
\thanks{Corresponding Author: Senior Quantum Applications Engineer, Lockheed Martin Aeronautics, 1 Lockheed Boulevard, Fort Worth, Texas, United States; E-mail: kristen.l.pudenz@lmco.com \\ Copyright 2016, Lockheed Martin Corporation. All rights reserved.}},
\author{\fnms{Gregory S} \snm{Tallant}},
\author{\fnms{Todd R} \snm{Belote}},
and
\author{\fnms{Steven H} \snm{Adachi}}

\runningauthor{K. L. Pudenz et al.}
\address{Lockheed Martin, United States}

\begin{abstract}
The utility of satisfiability (SAT) as an application focused hard computational problem is well established. We explore the potential of quantum annealing to enhance classical SAT solving, especially where sampling from the space of all possible solutions is of interest. We address the formulation of SAT problems to make them suitable for commercial quantum annealers, practical concerns in their implementation, and how the performance of the resulting quantum solver compares to and complements classical SAT solvers.
\end{abstract}

\end{frontmatter}
\pagestyle{empty}

\section{Introduction}
Boolean satisfiability is a well known NP-complete problem that serves as the basis for many computationally hard tasks. Some of these applications, such as formal software verification and data mining, are well served by solvers which are capable of finding more than one satisfying assignment of variables, or even all such solutions \cite{toda2015implementing}. Quantum annealing (QA) has been applied to the solution of weighted max 2-SAT problems \cite{santra2014max,bian2010teaching}, the construction of SAT filters \cite{douglass2015constructing}, SAT variants \cite{roy2016mapping}, and frustrated loop problems inspired by SAT \cite{hen2015probing,king2015performance}. QA hardware operates by performing many annealing (solving) cycles, returning a potential solution sample for each cycle \cite{johnson2011quantum,bian2010teaching}. We seek to discover whether quantum annealers could provide an advantage in finding multiple solutions to satisfiability problems.

We first provide a construction for posing satisfiability problems to quantum annealing hardware solvers in Section \ref{sec_sat_penalty}. Section \ref{sec_mixedsat} defines our problem set, and Section \ref{sec_results} details the results of our quantum and classical solution sampling comparision.

\section{Cascading-OR SAT penalty functions}
\label{sec_sat_penalty}

Satisfiability problems in general consist of a conjunction of disjunctive clauses. The conjunction can be achieved on quantum annealing hardware by summing terms representing individual clauses, so we will focus here on defining the disjunctive clause Hamiltonian penalty functions, which we denote $H_k$, with $k$ the number of variables in each clause. The clauses are built using a cascading-OR construction, which requires $2(k-1)$ qubits per clause. It should be noted that the resulting clauses are not directly embeddable on QA hardware graphs. No direct embeddings are proposed here, primarily due to the fact that many clauses share variables, and by the time this is taken into account in building the Hamiltonian, a heuristic embedding algorithm for the problem as a whole will still be required. The overhead of heuristic embedders for QA hardware graphs is of course an entire line of research on its own.

\begin{figure}
\centering
\begin{subfigure}{0.45\textwidth}
	\includegraphics[width=0.5\textwidth]{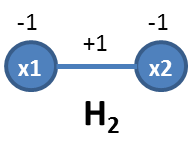}
	\caption{2-SAT penalty}
\end{subfigure}
\begin{subfigure}{0.45\textwidth}
	\includegraphics[width=0.5\textwidth]{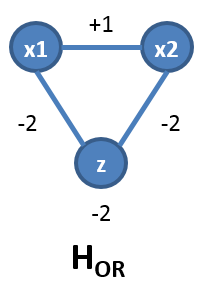}
	\caption{OR with output penalty}
\end{subfigure}
\caption{Building blocks for cascading-OR penalty functions.}
\label{blocks_figure}
\end{figure}

We begin our construction with two building blocks. The first is 
\begin{equation}
H_2(x1,x2) = -\sigma^z_{x1} - \sigma^z_{x2} + \sigma^z_{x1}\otimes\sigma^z_{x2} ,
\end{equation}
the ground states of which represent the states for which $x1\vee x2$ evaluates to true, i.e. the states for which the $2$-SAT clause is satisfied. We also define an "OR with output" penalty function
\begin{equation}
H_{OR}(x1,x2,z) = \sigma^z_{x1} + \sigma^z_{x2} - 2\sigma^z_z + \sigma^z_{x1}\otimes\sigma^z_{x2} - 2\sigma^z_{x1}\otimes\sigma^z_{z} - 2\sigma^z_{x2}\otimes\sigma^z_{z} ,
\end{equation}
which has as its ground states the states where  $x1 \vee x2 = z$, the crucial difference being that the correct state of the OR output is included in the penalty function. By substituting the output bit $z$ from the OR function for $x1$ in $H_2$, we can easily construct $H_3$:
\begin{align}
H_3 &= H_{OR}(x1,x2,z1) + H_2(z1,x3)\\
&= (\sigma^z_{x1} + \sigma^z_{x2} - 2\sigma^z_{z1} + \sigma^z_{x1}\otimes\sigma^z_{x2} - 2\sigma^z_{x1}\otimes\sigma^z_{z1} - 2\sigma^z_{x2}\otimes\sigma^z_{z1}) - \sigma^z_{z1}  - \sigma^z_{x3} + \sigma^z_{z1}\otimes\sigma^z_{x3} ,
\end{align}
where the part of the equation in parenthesis represents $H_{OR}$ applied to $x1$ and $x2$, with its output $z1$ cascaded into one of the qubits in $H_2$. Note that the term $\sigma^z_{z1}$ appears twice in the equation because it is shared between $H_{OR}$ and $H_2$, so the collected term would be $-3\sigma^z_{z1}$. See Figure \ref{blocks_figure} for diagrams of the basic penalty functions we have introduced.

\begin{figure}
\centering
\begin{subfigure}{0.45\textwidth}
	\includegraphics{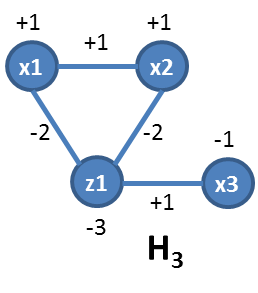}
	\caption{Option 1 for 3-SAT penalty}
\end{subfigure}
\begin{subfigure}{0.45\textwidth}
	\includegraphics{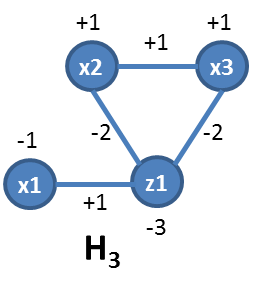}
	\caption{Option 2 for 3-SAT penalty}
\end{subfigure}
\begin{subfigure}{0.5\textwidth}
	\includegraphics{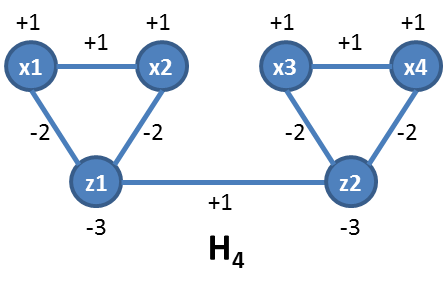}
	\caption{Option 1 for 4-SAT penalty}
\end{subfigure}
\begin{subfigure}{0.5\textwidth}
	\includegraphics{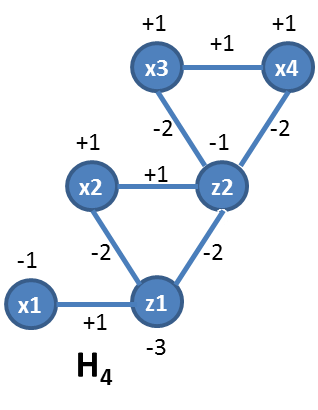}
	\caption{Option 2 for 4-SAT penalty}
\end{subfigure}
\caption{Options for $H_3$ and $H_4$. The penalty function can take different forms depending on which logical variable qubits are extended by $H_{OR}$; these forms become more varied as $k$ increases and the decision tree widens.}
\label{options_figure}
\end{figure}

Penalty functions $H_k$ of any size can be built inductively by extending this methodology. Start with a valid $H_{k-1}$, choose any of the qubits which directly represent problem variables (shown in above equations as $xi$), and substitute the output qubit of $H_{OR}$ into its place. The size of the penalty function grows by $2$ qubits because we are putting three in the place of one. Because we can choose any problem variable qubit, we can define multiple equivalent $H_k$ penalty functions depending on which substitution we use. Figure \ref{options_figure} shows several possible constructions of $H_3$ and $H_4$ penalty functions.

\section{Mixed SAT}
\label{sec_mixedsat}

The test problem set for this study consists of a group of mixed SAT problems which contain clauses of diverse lengths (as compared to the standard 3-SAT, in which each clause involves three variables). The test set was tailored to be feasible on the D-Wave 2X generation of quantum annealers. We ran $123$ instances with $n=30$ on a $1098$ qubit QA chip, taking $1000$ solution samples for each instance and parameter setting at an annealing time of $20$ microseconds per sample. These instances were a subset of a larger set of mixed SAT instances which was generated randomly, then downselected based on number of solutions (less than a million) and embeddability of the resulting penalty function on hardware.

\section{Results}
\label{sec_results}

\subsection{Timing metrics}
We considered two timing metrics for the quantum annealer results. The first, core annealing time, is important because it captures the key physics of the computation. The quantum annealer calculates the satisfying assignment by performing an adiabatic quantum evolution from a known original hardware state to the unknown problem solution. This process is the \textit{core anneal}, and is the key difference between quantum annealers and classical hardware. By separating the core anneal time from other hardware times, we get an idea of the essential limits on this type of computation.

We also present results using wallclock time for the quantum annealer. This includes, among many things, programming time, thermalization time (so the chip can cool down after programming), core annealing time, readout time, and postprocessing time. The wallclock time is how long we have to wait for results from the current generation of quantum annealers. However, there is reason to believe that many of the elements within the wallclock time will be subject to engineering improvement in the near term.

\subsection{Classical comparison}
We used a recently published classical ALL-SAT solver \cite{toda2015implementing} to count the solutions for each problem and establish a timing baseline. Toda and Soh programmed several ALL-SAT modifications to MiniSAT, a well known single-solution SAT solver. We used the version that performed best in their benchmarks. The optimality of this solver for ALL-SAT problems in general and mixed SAT in particular is not established, but it is the current state of the art. All times shown for the classical solver are wallclock times. The solver timestamped each solution, so we have granular timing data as to when each solution was found.

\subsection{Core annealing time}
\begin{figure}
\centering
\includegraphics[width=\textwidth]{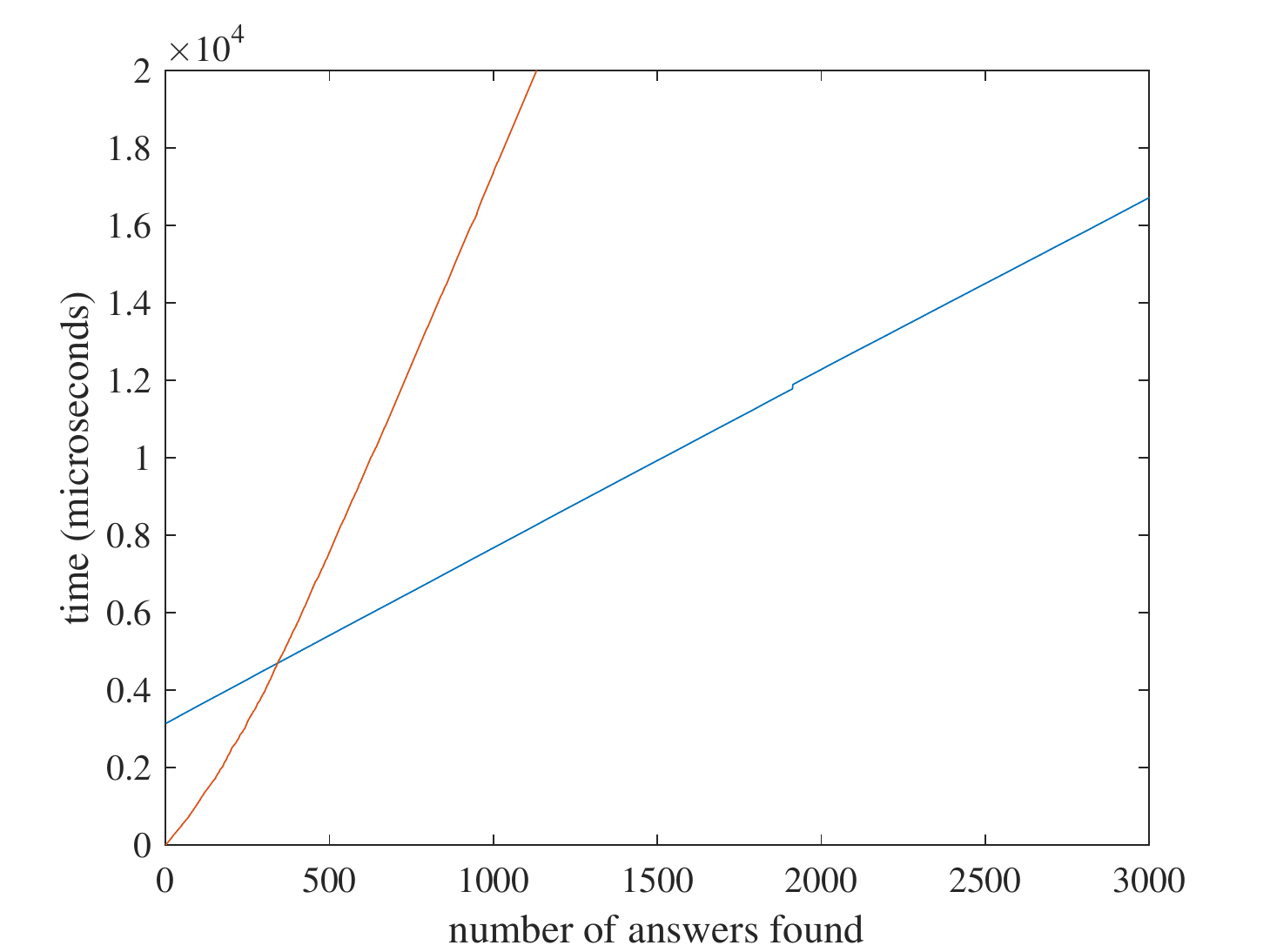}
\caption{Rate of solution finding for classical and quantum solvers. This plot is for an instance with one of the larger initial solution finding advantages compared to the classical alternative. The shapes of the curves are consistent across all instances in the problem set.}
\label{fig_instance_2_anneal}
\end{figure}

\begin{figure}
\centering
\includegraphics[width=\textwidth]{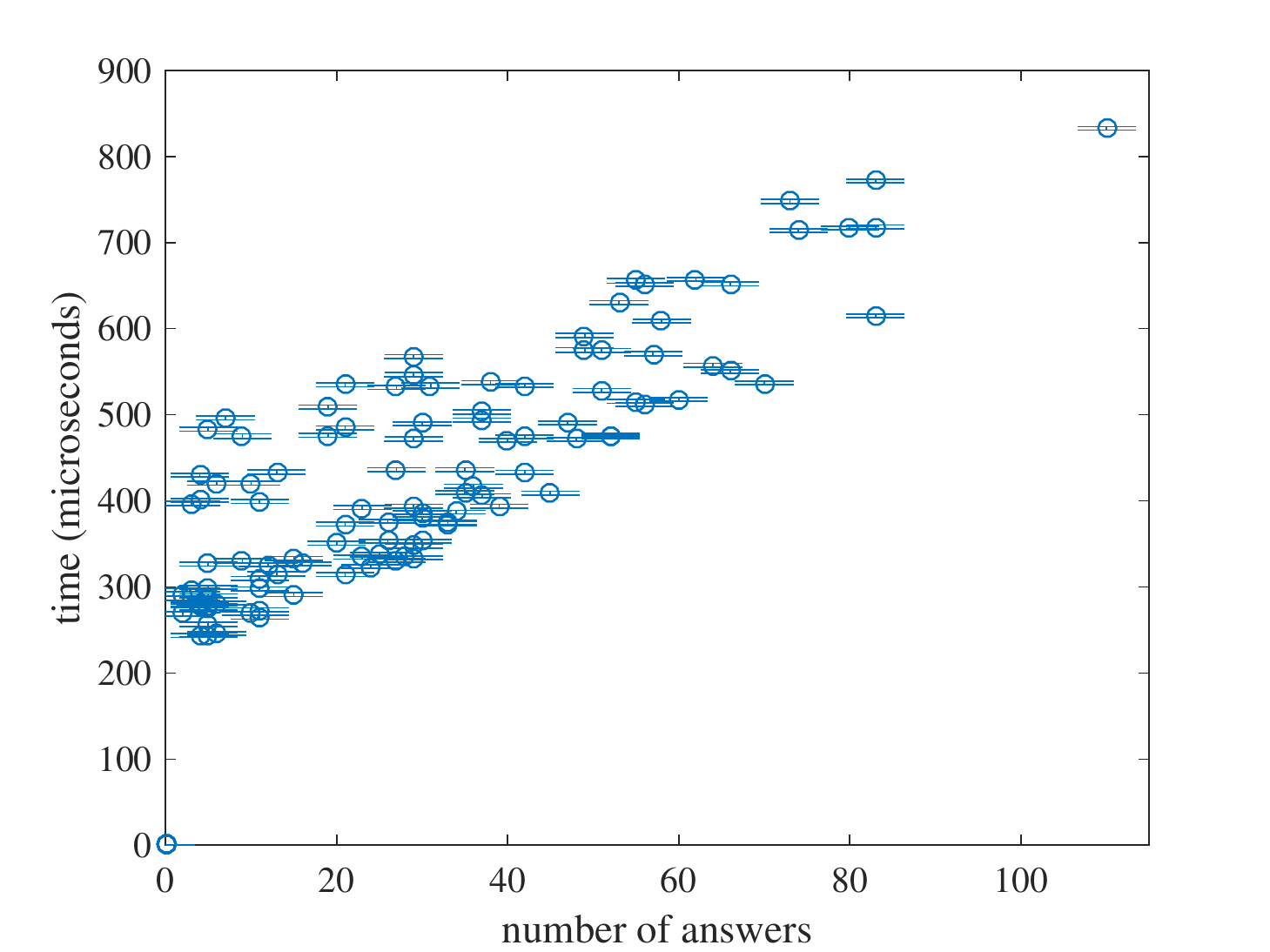}
\caption{Crossover points between classical and quantum solution finding rates. Each point represents one problem instance and shows the location of the crossover we see for a single problem in Figure \ref{fig_instance_2_anneal}. Error bars represent the resolution of the timing data.}
\label{fig_crossover}
\end{figure}

\begin{figure}
\centering
\includegraphics[width=\textwidth]{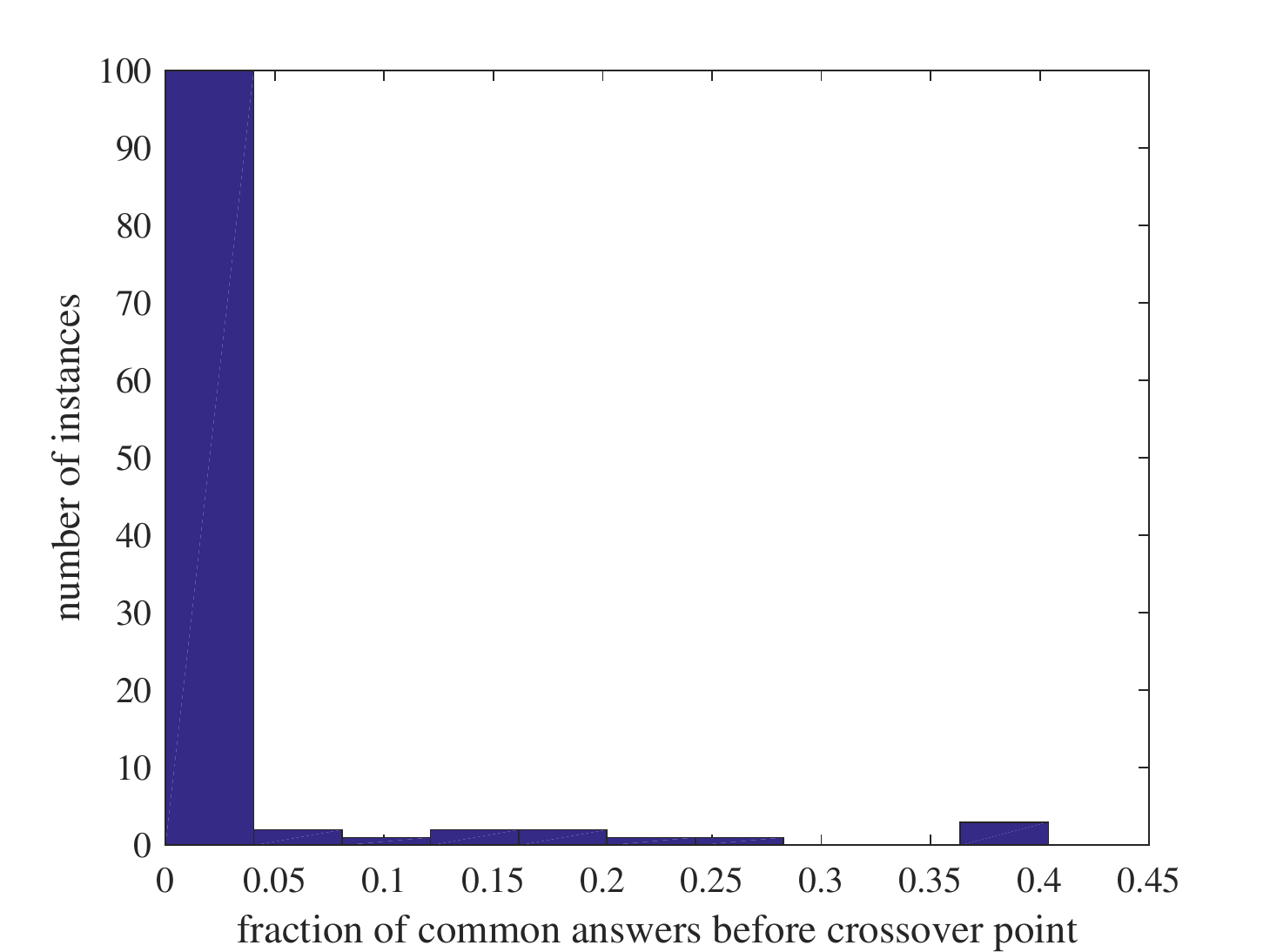}
\caption{Overlap between classical and quantum solution sets at crossover. Bars indicate the number of problem instances with a given fraction of solutions appearing in both the classical and quantum partial solution sets at the crossover point.}
\label{fig_overlap}
\end{figure}

When only core annealing time is considered, most instances in the problem set initially find solutions faster than the classical solver. In Figure \ref{fig_instance_2_anneal}, the classical and quantum curves for time to find distinct solutions are shown. The quantum annealer initially finds solutions much more quickly, but slows in the rate at which it finds new solutions, allowing the classical curve to cross under it. Figure \ref{fig_crossover} shows these crossover points for the entire instance set. We find an initial quantum advantage in the tens of solutions for most instances. It is relevant to note that the sets of solutions found by the classical and quantum solvers at the crossover point have low overlap (see Figure \ref{fig_overlap} for details).

\subsection{Wallclock time}
\begin{figure}
\centering
\includegraphics[width=\textwidth]{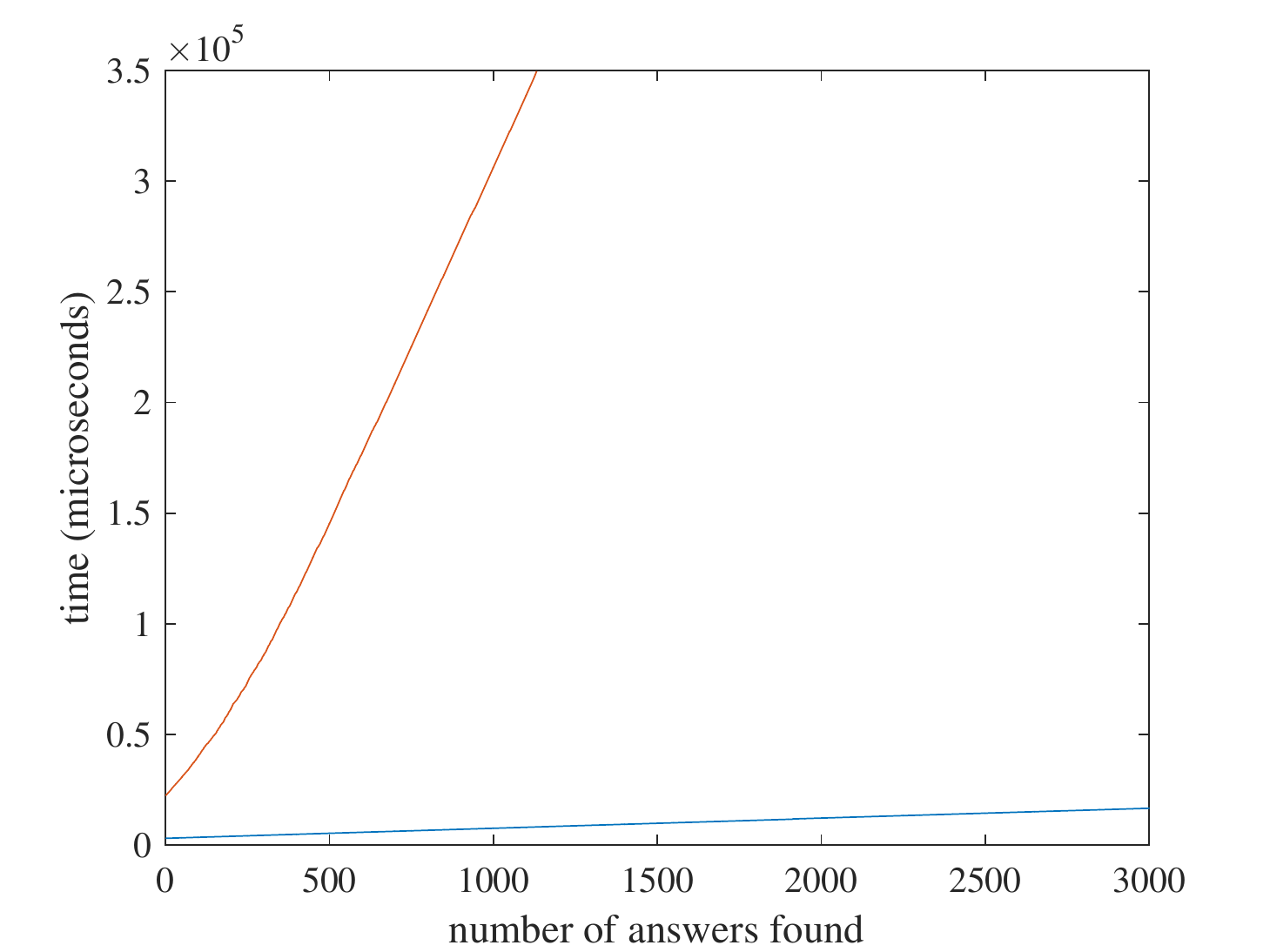}
\caption{Wallclock rate of solution finding for classical and quantum solvers. This is the same problem instance shown in Figure \ref{fig_instance_2_anneal}, with the quantum curve measured in wallclock rather than core annealing time. Any quantum advantage is lost.}
\label{fig_instance_2_wallclock}
\end{figure}

When all quantum processor activities are accounted for, and we compare quantum wallclock to classical wallclock time, the initial advantage disappears. Figure \ref{fig_instance_2_wallclock} shows the same instance as Figure \ref{fig_instance_2_anneal}, but with the quantum wallclock rather than core anneal time. The quantum curve is immediately at a disadvantage and never crosses the lower classical time curve. This is true for all instances in the set we studied.

\subsection{Solution sampling}
\begin{figure}
\centering
\includegraphics[width=\textwidth]{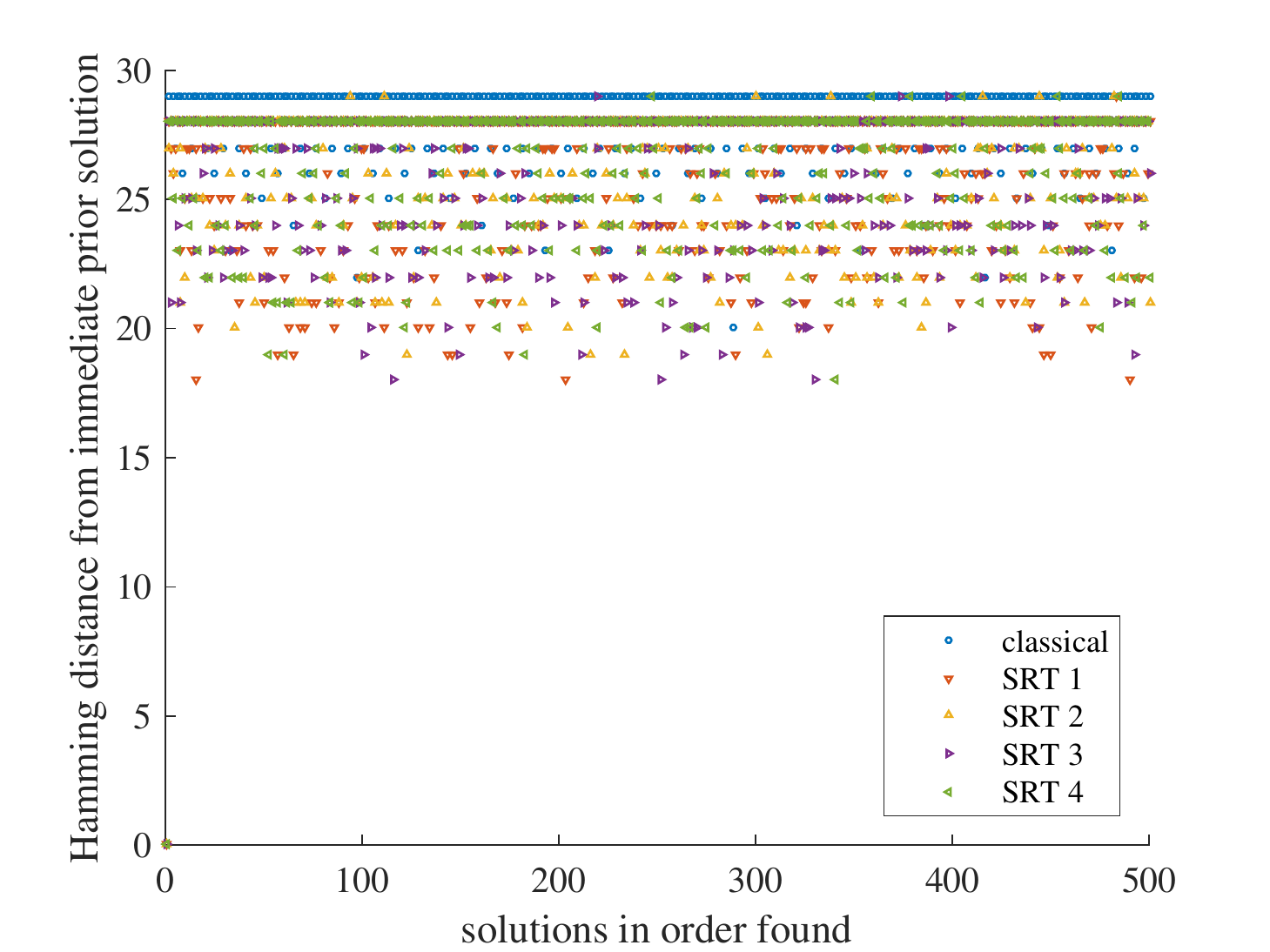}
\caption{Hamming distance between neighboring solutions. The solutions for a typical problem instance are displayed here in the order they were returned from each solver. Each point represents the Hamming distance of the current solution from the one that came before it. Blue circles represent solutions from the classical solver. All other colors and shapes represent solutions from the quantum annealer, each type of marker indicating one trivial spin reversal transformation (SRT) of the problem's penalty function. All spin reversals on the quantum annealer behave consistently, but the distance between adjacent classical solutions is both higher and more consistent than the quantum solver can produce.}
\label{fig_hamming_neighbor}
\end{figure}

Quantum annealers solve problems by sampling from the low energy state space of the penalty function, so it is natural to consider the application of quantum annealers as solution samplers for problems of interest. In particular, we were interested in whether quantum annealers could provide more variety in the answer set than classical solvers, particularly because the quantum penalty function can be subjected to trivial spin reversal transformations or parameter settings that leave the identity of the ground state solutions the same but change the physics of the solving process \cite{ronnow2014defining,pudenz2016parameter}. To measure the diversity of the solution set, we look at the Hamming distance between neighboring solution samples (Figure \ref{fig_hamming_neighbor}). We find that the classical solver is both more consistent in the distribution of solutions and finds solutions that are more widely separated in Hamming distance.

\section{Conclusions}
Quantum annealers have the potential to assist in finding multiple solutions to mixed satisfiability problems, but in their current form are not ready for the task. The advantage in time for finding an initial set of solution samples seen when only core annealing time is considered and the low overlap between the quantum and classical crossover solution sets points to a potential future role for a quantum solver as a helper to a classical solver. Indeed, it might be possible to run the quantum solver to crossover with the classical solver, apply a spin reversal transformation, and run the quantum solver again to find another set of solutions. However, with current quantum annealing technology, the overhead timing factors included in the wallclock time destroy any advantage.

\end{document}